\documentclass[journal]{IEEEtran}
\usepackage{bm}
\usepackage{graphicx}
\usepackage{caption}
\usepackage{mathrsfs}
\usepackage{amsmath}
\usepackage{array}
\usepackage[table]{xcolor}
\usepackage{color}
\usepackage{epsfig,epstopdf,subfigure}
\usepackage{subfigure}

\begin{document}

\title{Deep Normalization for Speaker Vectors}

\author{Yunqi Cai, Lantian Li, Dong Wang and Andrew Abel
\thanks{

This work was supported by the National Natural Science Foundation of China (NSFC) under the project No.61633013 and No.61371136.
Thanks to Dr. Zhiyuan Tang's valuable discussion. Dong Wang is the corresponding author. Y. Cai and L. Li are joint first authors.

Y. Cai is with the Center for Speech and Language Technologies (CSLT) and the Department of Computer Science
at Tsinghua University, Beijing 100084, China (e-mail: caiyq@cslt.org).

L. Li and D. Wang are with the Center for Speech and Language Technologies (CSLT), BNRist
at Tsinghua University, Beijing 100084, China (e-mail: wangdong99@mails.tsinghua.edu.cn, lilt@cslt.org).

A. Abel is with the Department of Computer Science and Software
Engineering, Xi'an Jiaotong-Liverpool University, Suzhou 215123, China (e-mail: andrew.abel@xjtlu.edu.cn).

}

}

\maketitle

\begin{abstract}

Deep speaker embedding has demonstrated state-of-the-art performance in speaker recognition tasks. However, one potential issue with this approach is that the speaker vectors derived from deep embedding models tend to be non-Gaussian for each individual speaker, and non-homogeneous for distributions of different speakers. These irregular distributions can seriously impact speaker recognition performance, especially with the popular PLDA scoring method, which assumes homogeneous Gaussian distribution.  In this paper, we argue that deep speaker vectors require deep normalization, and propose a deep normalization approach based on a novel discriminative normalization flow (DNF) model. We demonstrate the effectiveness of the proposed approach with experiments using the widely used SITW and CNCeleb corpora. In these experiments, the DNF-based normalization delivered substantial performance gains and also showed strong generalization capability in out-of-domain tests.

\end{abstract}


%
\IEEEpeerreviewmaketitle

\section{Introduction}
\label{sec:intro}

Speaker recognition is widely studied, and decades of investigation has resulted in significant performance improvements, and deployment in a wide range of practical applications~\cite{campbell1997speaker,reynolds2002overview,hansen2015speaker}.  Traditional speaker recognition methods are based on statistical models, which include the popular Gaussian mixture model-universal background model (GMM-UBM) architecture~\cite{Reynolds00}. In order to improve the statistical strength with limited data, various subspace models have been proposed~\cite{Kenny07}, and in particular, the i-vector model, which was the most successful~\cite{dehak2011front}, especially accompanied with a
probabilistic linear discriminant analysis (PLDA)~\cite{Ioffe06} scoring model.


In recent years, deep learning methods have demonstrated significant progress with regard to speaker recognition fields. Variani et al. reported the results of an initial investigation on a text-dependent task~\cite{ehsan14}, using a deep neural net (DNN) to produce frame-level speaker-discriminant features.  They then derived utterance-level representations (called `\textbf{d-vectors}') by average pooling.  Li et al.~\cite{li2017deep} extended this with a more speech-friendly net structure, and achieved good performance on text-independent tasks, especially with short utterances. However, one key shortcoming of the early d-vector approach is that the frame-level training does not match the utterance-level test.

Researchers developed two architectures to solve this problem.
The first approach is to use the \emph{end-to-end architecture}, which accepts two utterances and produces the accept/rejection decision directly~\cite{heigold2016end,zhang2016end,zhang2017end,Chowdhury17attention,lic2017deep,zhang2018text}.  The second approach is the \emph{deep speaker embedding architecture}~\cite{snyder2018xvector,okabe2018attentive}, which instead accumulates the frame-level statistics of a variable-length utterance and converts these to a fixed-length vector with the objective of discriminating between the speakers in the training set.  Although the training criterion of the end-to-end approach is more consistent with the speaker recognition task, the deep embedding approach is easier to train~\cite{wang2017deep} and the derived speaker vectors support various speaker-related tasks such as speaker-dependent synthesis~\cite{jia2018transfer}.
Perhaps the most popular deep embedding architecture is the x-vector model proposed by Snyder et al.~\cite{snyder2018xvector}.  It is based on a time-delayed neural net and a statistical pooling, and has achieved good performance in many applications.  For simplicity, we use the term \textbf{x-vector} to represent deep speaker vectors derived with any net structure.\footnote{In this paper, we use \emph{speaker embedding} and \emph{speaker vector} to refer to the embedding technique and the representations produced by this technique, respectively, though some researchers use both of them to
refer the representations.}

Recently, deep speaker embedding models have been significantly improved by developing a more comprehensive architecture~\cite{chung2018voxceleb2,Jung2019raw}, improved pooling methods~\cite{okabe2018attentive,Cai2018,Xie19a,Chen2019tied}, training criteria~\cite{li2016max,ding2018mtgan,Wang2019centroid,bai2019partial,Gao2019improving,Zhou2019deep}, and training schemes~\cite{Li2019boundary,Wang2019phonetic,Stafylakis2019}. As a result, the deep embedding approach has achieved state-of-the-art (SOTA) performance~\cite{Sadjadi2019}.

\begin{figure}[htb]
    \centering
    \includegraphics[width=0.8\linewidth]{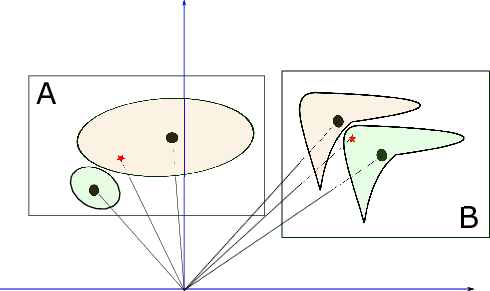}
    \caption{Illustration of potential problems caused by non-Gaussian and non-homogeneous distributions. Each colored region represents the distribution of a particular speaker, and the region boundary represents the contour of the same probability.  For simplicity, the classification is based on cosine distance. (A) shows two non-homogeneous distributions (although both are Gaussian). The test utterance (red star) is categorized as being the cyan speaker, although the probability that it belongs to the brown speaker is higher. (B) shows two non-Gaussian distributions (although they are homogeneous). The test utterance (red star) is categorized as the brown speaker but the probability that it belongs to the cyan speaker is higher. }
    \label{fig:flaw}
\end{figure}

Despite the significant progress outlined above, one potential issue with the deep embedding approach is that the training objective of the embedding models is purely discriminative, meaning that the goal of the training is simply discriminating the speakers, without considering the distribution of the derived speaker vectors. As a consequence, the produced speaker vectors tend to be
irregular.
This means that:  (1) the distributions of each individual speaker may be potentially very complex and far from Gaussian (\textbf{non-Gaussianality}), and (2) the distributions of different speakers may be significantly different (i.e. \textbf{non-homogeneity}).  Non-Gaussian and non-homogeneous distributions may seriously impact performance of the back-end scoring model, particularly with regard to the popular PLDA scoring method, which is based on the assumption that all speaker distributions are homogeneous and Gaussian.  Fig.~\ref{fig:flaw} illustrates some potential problems caused by non-Gaussian and non-homogeneous distributions.  It should also be noted that this issue is not as severe for statistical speaker vectors (i-vectors) as they are derived from a constrained model with an underlying Gaussian assumption.

A number of researchers have noticed the risk associated with data irregularity, and have presented various compensations.  For instance, phone-aware training was investigated, with the aim of reducing the non-Gaussian variation caused by speech content~\cite{li2015improved,Wang2019phonetic}.  Another proposed approach was to use various data augmentation methods~\cite{snyder2018xvector,Wu2019data}.
These augmentation methods prevent the training from being over-fitted to highly curved discrimination boundaries, hence more regulated distribution for each individual speaker. Li et al. presented an approach that treats the full-connection layer before the softmax classifier as the basis of speakers, and imposes a central loss~\cite{li2018full,li2019gaussian}. This central loss is an explicit regularization that encourages the distributions of individual speakers to be more Gaussian. This concept was also introduced by other researchers~\cite{Cai2018}.


Previous work by the authors presented a variational auto-encoder (VAE) based normalization for x-vectors~\cite{zhang2019vae,wang2019vae}.  This is an independent component and dedicated to regulating x-vectors, and differs from other previous research that integrates the normalization constraint in the embedding model.  Our VAE-based normalization encourages the \emph{marginal distribution} to be Gaussian. However, it cannot normalize \emph{conditional distributions}, i.e. the distributions of individual speakers, a more important source of data irregulation. In addition, to retain the discriminant strength of the speaker vectors, an auxiliary cross-entropy loss is required when training the VAE model, which complicates the behavior of the normalizer.

In this paper, we propose a new fully generative model to normalize deep speaker vectors. This model is based on normalization flow (NF), a simple yet powerful architecture for density estimation. With this model, a complex distribution can be transformed to a simple isotropic Gaussian (often called the prior distribution).  However, directly applying the NF model is insufficient: it regulates the marginal distribution rather than conditional distributions (as with VAE), so cannot deal with individual speakers. We therefore propose a novel discriminative normalization flow (DNF). Compared with the vanilla NF model, DNF allows class-specific prior distributions, which enables it to model multiple speakers with different but homogeneous isotropic Gaussians. This paper will show that our new DNF approach is a deep and nonlinear extension of the widely used linear discriminant analysis (LDA) model.



The remainder of the paper is organized as follows. We first review the LDA and PLDA model in Section~\ref{sec:lda}, and through experiments, investigate the role of normalization that LDA plays when applied to x-vectors with PLDA scoring. Section~\ref{sec:dnf} presents the DNF model, which extends the shallow and linear normalization with LDA to a deep and and nonlinear normalization, with experimental results and analysis presented in Section~\ref{sec:exp}, and finally, the paper is concluded in Section~\ref{sec:con}.

\section{Shallow normalization by LDA}
\label{sec:lda}

It is well known that for x-vector systems with PLDA scoring, LDA is an important pre-processing step. This is initially unexpected, as PLDA is theoretically an extension of LDA, and can self-discover the most discriminant dimensions. Previous work~\cite{zhang2019vae,li2019gaussian} by the authors argued that the role LDA plays is distribution normalization, by making the conditional distributions of speaker vectors more Gaussian. This normalization makes the LDA-projected data fit the assumptions of PLDA, and therefore makes it more suitable for PLDA modeling.  However, this argument should be investigated in more depth, with more theoretical and empirical analysis.

\subsection{Review of LDA and PLDA}

\subsubsection{Dimension reduction view for LDA}

LDA is a popular tool for dimension reduction, by identifying the most class-discriminant directions, along which the between-class variation is maximized and the within-class variation is minimized (Fisher criterion)~\cite{fisher1936use}.  These directions are aligned with the eigenvectors of $\mathbf{S}_B \mathbf{S}^{-1}_W$ with large eigenvalues, where $\mathbf{S}_B$ and $\mathbf{S}_W$ are the between-class and within-class covariance matrices, respectively. An important property of these eigenvectors is that they diagonalize $\mathbf{S}_B$ and $\mathbf{S}_W$ \emph{simultaneously}.

Research has shown that LDA can be performed in two steps: normalization and discrimination~\cite{hastie2009elements}.  Normalization is a linear transform where the within-class covariance becomes an identity matrix $\mathbf{I}$. This can be achieved with principal component analysis (PCA) on the mean-offset data, and a re-scaling operation.  The discrimination is an extra orthogonal transform that aligns the data coordinates along the directions with the largest between-class variation. This can be achieved by another PCA on the normalized class means, and dimension reduction can be achieved by selecting the leading principal components (PCs). Within this discrimination space, classification will be optimal based on Euclidian distance, as shown in Fig.~\ref{fig:lda}.

\subsubsection{Probabilistic model view for LDA}

\begin{figure}[htbp]
\centering\includegraphics[width=\linewidth]{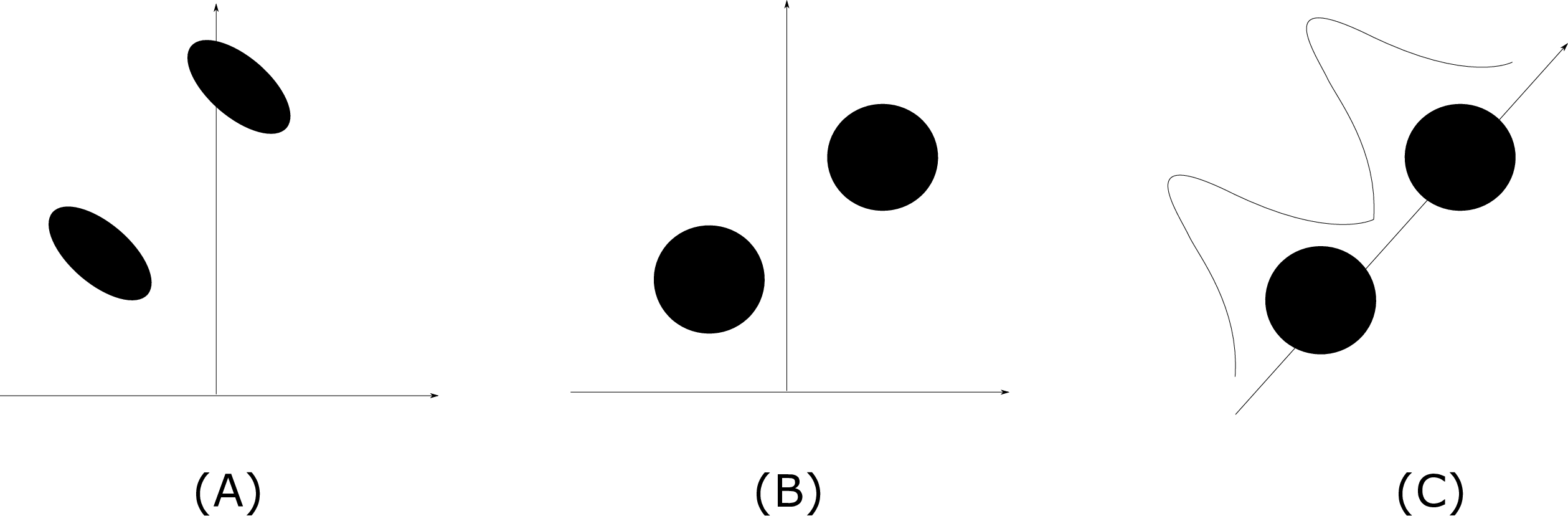}
\caption{The multi-Gaussian view of LDA.
(A) In the primary form, both classes are Gaussian and share the same covariance. (B) Normalization step: To discover the most discriminant features, firstly transform conditional distributions to be isotropical Gaussian, so that MAP classification can be performed by Euclidian distance.
(C) Linearly transform class means to a subspace with the largest between-class variation. This is an orthogonal transform, so does not change the shape of the within-class covariance. Therefore, Euclidian distance based classification remains optimal in terms of MAP prediction.
}
\label{fig:lda}
\end{figure}

LDA can be cast to a multi-Gaussian model~\cite{hastie1996discriminant}. Assuming all classes are Gaussian distributed and share the same covariance and the priors of all classes are equal, a probabilistic model can be established. The parameters, including the mean vectors $\{\mathbf{z}_c\}$ and the shared covariance matrix $\pmb{\Sigma}$, can be optimized following the maximum-likelihood (ML) criterion. After training, optimal classification (in terms of maximum a posterior, MAP) for a sample $\mathbf{x}$ can be obtained by choosing the class $c$ with the largest $p(\mathbf{x}|c)$. This is the primary form of LDA. The dimension reduction can be recovered by a constrained multi-Gaussian model where the mean vectors of all the classes are in a subspace.  Fig.~\ref{fig:lda} illustrates the multi-Gaussian view of LDA and its relationship with the dimension-reduction view.

As discussed, LDA assumes that the within-class distributions are Gaussian and they share the same covariance. We denote these as the \textbf{Gaussianality condition} and the \textbf{homogeneity condition}, respectively.  If both conditions are satisfied, we call the data \textbf{regulated}. Regulated data are suitable for LDA dimension reduction, otherwise the features selected would not be optimal in terms of class discrimination.

\subsubsection{PLDA}

PLDA extends LDA (probabilistic model view), by placing a Gaussian prior on the class means. A key advantage associated with this prior is that it enables dealing with new classes. More specifically, the posterior distribution of the mean of a new class can be derived, given even a single sample. According to this posterior, the probability that one or more test samples belong to the new class can be calculated by marginalizing over the class mean. Therefore, PLDA can be used to compute the likelihood ratio that two utterances belong to the same and different speakers~\cite{Ioffe06}, and is a theoretically sound scoring model. It is important to highlight that PLDA inherits the shared Gaussian assumption of LDA and requires regulated data. If the data are irregulated, the likelihood ratio derived by PLDA may be biased.

\subsection{Why LDA works for x-vectors}

To have a better understanding of the role of LDA and explain its contribution to the back-end scoring model, in particular PLDA,
we performed an initial experiment using a subset of the VoxCeleb dataset~\cite{chung2018voxceleb2,nagrani2017voxceleb}.
We firstly created both the i-vector and x-vector systems using the entire VoxCeleb database,
following the standard Kaldi SITW~\cite{povey2011kaldi} recipes.
We then generated i-vectors and x-vectors from a subset of 600 speakers in the training set. We use these to investigate the statistical properties of these vectors before and after LDA.
More details about the experimental setup will be described in Section~\ref{sec:exp}.

\subsubsection{Global properties}

\begin{figure}[htbp]
\centering\includegraphics[width=0.4\linewidth]{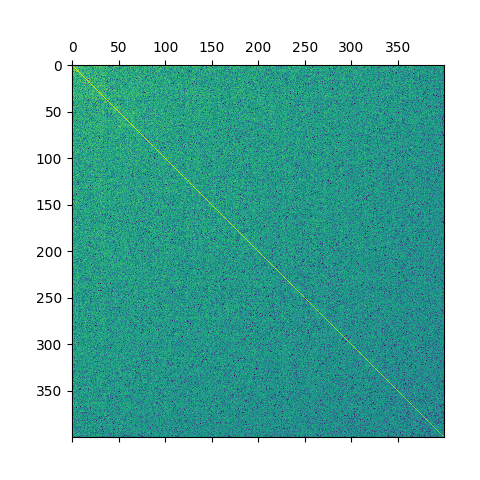}
\centering\includegraphics[width=0.4\linewidth]{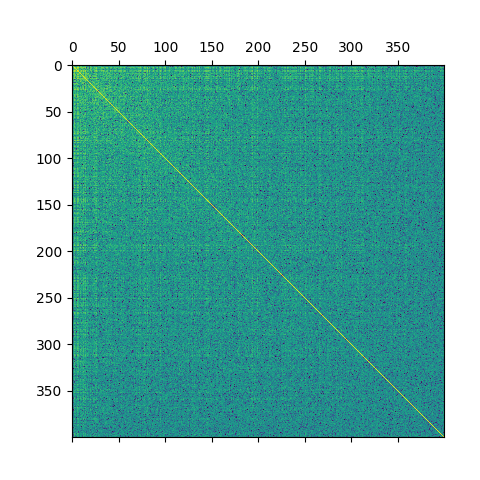}
\centering\includegraphics[width=0.4\linewidth]{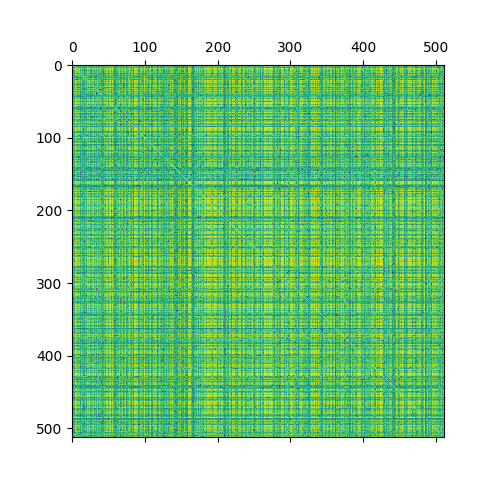}
\centering\includegraphics[width=0.4\linewidth]{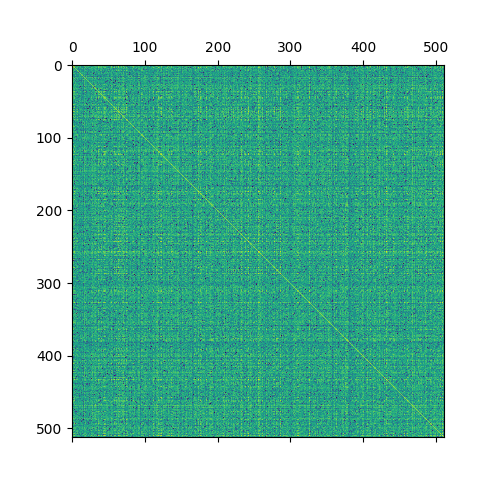}
\caption{Between-speaker (left) and within-speaker (right) covariance of i-vectors (top) and x-vectors (bottom).}
\label{fig:cov}
\end{figure}

First of all, we show the between-speaker and within-speaker covariance matrices of i-vectors and x-vectors.  These matrices reflect the global properties (i.e. on the data of all the speakers) of the distribution of the speaker vectors. As shown in Fig.~\ref{fig:cov}, x-vectors exhibit more complex correlation patterns compared to i-vectors.  After LDA, for both x-vectors and i-vectors, the between- and within-speaker covariances become diagonal.  This diagonalization can be regarded as a global normalization. As the correlation patterns of x-vectors are more complex, this normalization is more substantial for x-vectors.  To an extent, it answers the question of why LDA contributes more for x-vectors with cosine scoring:  although x-vectors are derived from a discriminative model and LDA-based feature selection seems less important, the diagonal between- and within-speaker covariances are good for cosine scoring (which assumes that the feature dimensions are independent).

\subsubsection{Local properties}
\label{sec:metric}

The global normalization of LDA does not fully explain everything.  In particular, it cannot explain why LDA contributes to PLDA scoring, as PLDA performs the same normalization anyway.  Our hypothesis is that LDA, by discarding less discriminative dimensions, performs a local normalization at the class level.  More precisely, we will show that LDA improves both homogeneity among classes and also the Gaussianality of each class.

We test this hypothesis with both i-vectors and x-vectors. The tests are conducted in three spaces: (1) the original observation space; (2) the LDA space, i.e. the subspace with leading discriminant dimensions after LDA transform; (3) the residual space, i.e. the subspace complementary to the LDA space. Since testing the homogeneity and Gaussianality in a high-dimensional space is challenging, we perform tests on the principal directions of each conditional distribution. Specifically, we select speakers with more than $100$ samples and perform PCA on the data of each speaker, and investigate the homogeneity and Gaussianality on the leading PC directions. The statistics we collected are as follows:

\begin{itemize}
\item PC direction variance for homogeneity. This tests if the covariance matrices of all the speakers have the same PC directions.  After PCA, the first PC (PC1) of all the speakers are selected and its mean over the speakers is computed. The cosine distance between the PC1s of individual speakers and the mean PC1 is computed. The variance of these cosine scores is used as the measure to test the PC1 direction variance. The same computation is conducted on all PCs. In this section, we report the direction variance on PC1 and PC2, and the averaged direction variance on the first 10 PCs.

\item PC shape variance for homogeneity. Using PC1 as an example, the coefficients (eigenvalues) of the covariance matrices of all the speakers on the first PC are calculated, and the variance of these coefficients over all speakers is computed. The same computation is performed on all the PCs.  Since the coefficient on each PC determines the spreading of the samplings on this direction, the coefficients on all the PCs determine the shape of the speaker distribution. The variances of these coefficients over all speakers then test if the distributions of all speakers have the same shape (regardless of the directions), hence being noted as PC shape variances. We report the shape variance on PC1 and PC2, and the averaged shape variance on the first 10 PCs.

\item Average PC kurtosis for Gaussianality. On each PC direction, we compute the kurtosis for each speaker, and then compute the mean of the kurtosis over all the speakers. The averaged kurtosis over the first 10 PCs is reported.

\item Average PC skewness for Gaussianality. On each PC direction, we compute the skewness for each speaker, and then compute the mean of the kurtosis over all the speakers. The averaged skewness over the first 10 PCs is reported.

\end{itemize}

The results are shown in Table~\ref{tab:var}, where we also report the between- and within-speaker variance, as well as the evaluation performance with the cosine scoring in terms of equal error rates (EER). There are several key observations:

\begin{enumerate}

\item Comparing i-vectors and x-vectors in the original observation space, it can be seen that x-vectors possess larger PC direction and PC shape variances, demonstrating that the distributions of different speakers are less homogeneous. Moreover, x-vectors show larger kurtosis and skewness, indicating that the distributions of individual speakers are less Gaussian. This confirmed our conjecture that x-vectors are less regulated, and are less suitable for PLDA scoring when compared to i-vectors.



\item For both i-vectors and x-vectors, after LDA, homogeneity and Gaussianality are all improved, and this improvement is much more significant for x-vectors. This indicates that LDA-transformed data are more regularized, especially for x-vectors.

\item In the residual space, LDA improves homogeneity. With regard to Gaussianality, there is a slight improvement with i-vectors. For x-vectors, however, Gaussianality becomes worse after LDA.

\end{enumerate}

\begin{table}[htb!]
 \begin{center}
  \caption{Homogenity and Gaussianality of i-vectors and x-vectors before and after LDA.}
  \label{tab:var}
   \begin{tabular}{|l|c|c|c|c|c|c|}
   \hline
                 & \multicolumn{3}{c|}{i-vector} & \multicolumn{3}{c|} {x-vector}\\
   \hline
                 & Orig. & LDA  & Res. & Org. & LDA & Res.\\
   \hline
   EER\%    & 5.75   &  3.08   &  19.83  & 7.00  & 1.50 & 15.08 \\
   \hline
   PC1 dir. var       & 0.064  & 0.009 & 0.004 & 0.104 & 0.006 & 0.003 \\
   PC2 dir. var       & 0.089  & 0.008 & 0.005 & 0.156 & 0.005 & 0.003 \\
   Avg PC dir. var    & 0.028  & 0.007 & 0.004 & 0.060 & 0.006 & 0.003 \\
   \hline
   PC1 shape var   &80.1   & 64.0 & 60.5 &  156.0 & 42.0  & 128.0\\
   PC2 shape var   &53.3   & 32.3 & 34.6 &  68.3  & 21.8  & 63.0 \\
   Avg PC shape var&30.7   & 19.8 & 20.9 &  42.0  & 13.7  & 32.5 \\
   \hline
   PC Kertosis       &1.579  &0.734 & 1.268 &  2.615 & 1.686 & 31.40 \\
   PC Skewness       & 0.311 & 0.209 & 0.309 & 0.369  & 0.275 & 1.110  \\
   \hline
   Between-class var &0.269 & 1.164 &0.163 & 0.548 & 2.332 & 0.225 \\
   Within-class var  &0.753 & 0.996 &0.991 & 0.192 & 1.030 & 1.026 \\
   \hline
  \end{tabular}
 \end{center}
\end{table}

\begin{figure}[htbp]
\centering\includegraphics[width=\linewidth]{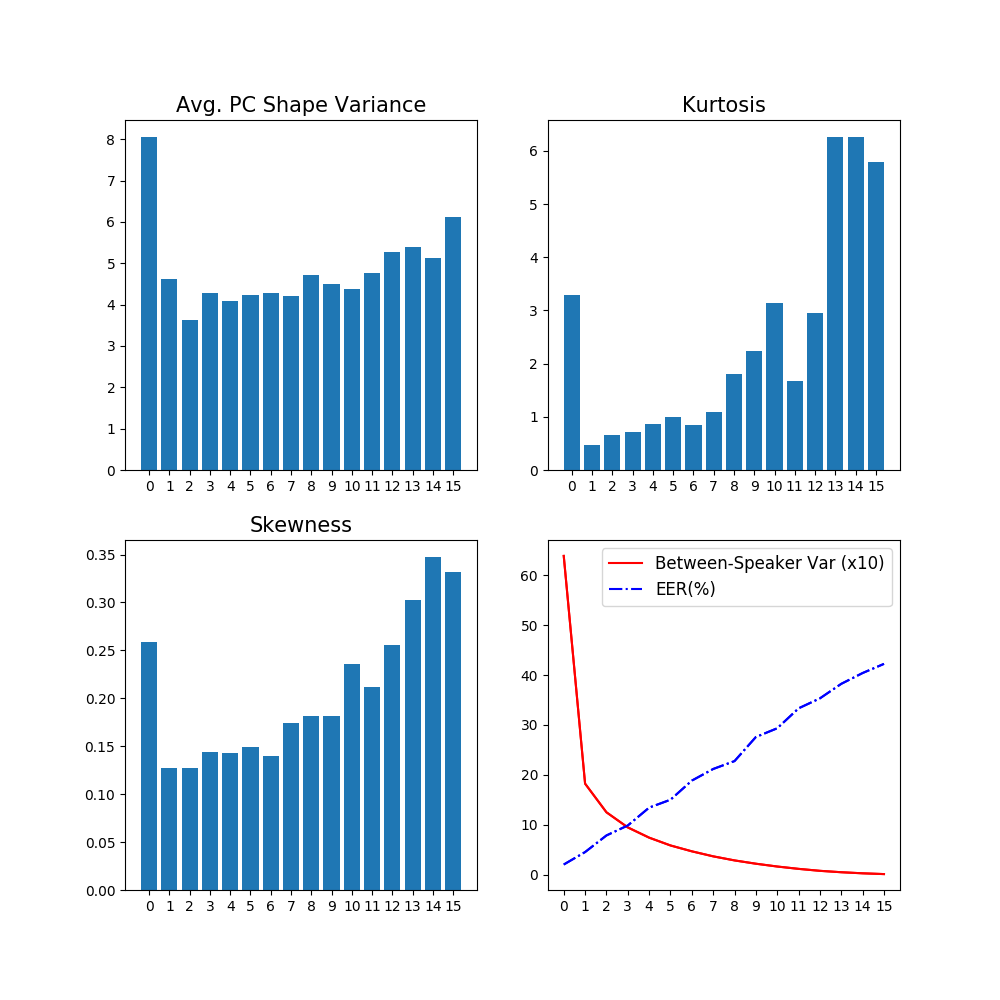}
\caption{Statistics of subgroup dimensions of LDA-projected x-vectors. Top left: Averaged PC shape variance; Top right: Kurtosis; Bottom left: Skewness; Bottom right: Between-speaker variance and EER with cosine scoring.}
\label{fig:spec-xv}
\end{figure}

\begin{figure}[htbp]
\centering\includegraphics[width=\linewidth]{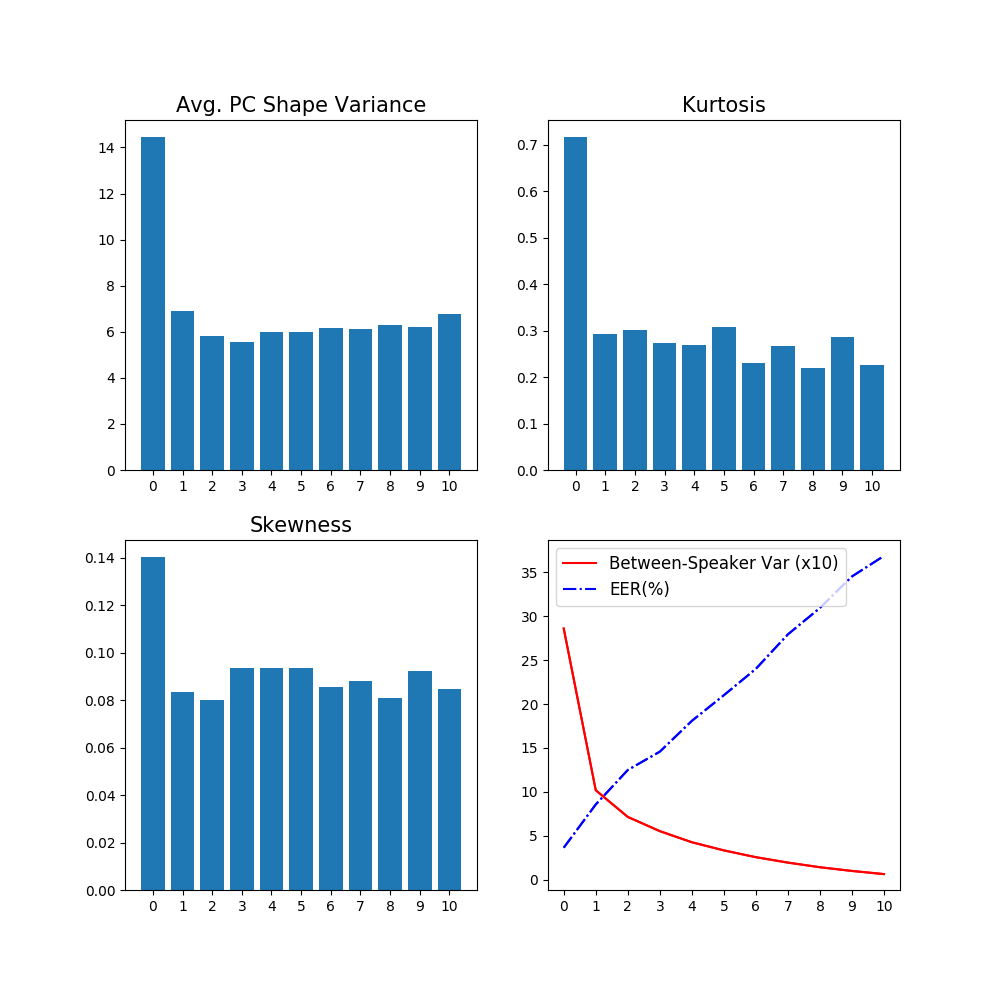}
\caption{Statistics of subgroup dimensions of LDA-projected i-vectors. Top left: Averaged PC shape variance; Top right: Kurtosis; Bottom left: Skewness; Bottom right: Between-speaker variance and EER with cosine scoring.}
\label{fig:spec-iv}
\end{figure}

The final observation is of greatest interest.  This suggests that LDA not only selects discriminant dimensions, but also removes non-Gaussian dimensions. To test this conjecture in a more concrete way, we divide all the dimensions sorted by LDA (according to their discriminant power) into multiple subgroups and compute the averaged PC shape variance, PC kurtosis and PC skewness, plus the between-speaker variance and the EER results with cosine scoring.  The results are shown in Fig.~\ref{fig:spec-xv}. It can be clearly seen that the most indiscriminative dimensions are non-homogeneous and non-Gaussian. In comparison, Fig.~\ref{fig:spec-iv} shows the results with i-vectors, where it can be seen that the indiscriminative dimensions of i-vectors are much more regulated.

\subsubsection{Summary of LDA and PLDA}

Based on the findings above, there are a number of key conclusions that can be drawn with regard to the role that LDA plays for i-vectors and x-vectors.  The typical role of LDA is two-fold, it transforms the between- and within- covariance to be diagonal, and it also selects the most discriminant dimensions. All of these functions contribute to cosine scoring. However, for PLDA scoring, these functions are performed implicitly and LDA pre-processing is therefore not usually necessary. This is the case with i-vectors, but with x-vectors, LDA plays an extra role -- \emph {speaker vector normalization} -- by removing irregulated (non-homogeneous and/or non-Gaussian) dimensions.  These irregulated dimensions may potentially be attributed to unwanted variance such as linguistic content and length variance, but can also be simply attributed to the unconstrained nature of deep embedding models.  Removing these dimensions will make the data more regulated and hence benefit PLDA scoring.

\section{Deep normalization by discriminative normalization flow}
\label{sec:dnf}

The previous section demonstrated that the suitable data regulation is very important for PLDA scoring. However, the linear form of LDA means that it can only normalize the global structure (within- and between- class covariances), rather than individual speakers.  The within-speaker normalization is essentially achieved by dimension reduction.  However, the dimensions that are least discriminant are not necessarily consistent with the most irregulated dimensions.  This can be clearly seen in Fig.~\ref{fig:spec-xv}, where the most non-Gaussian and non-homogeneous dimensions are in the first subgroup, i.e., the most discriminant dimensions. This means that dimension reduction is not optimal for either selecting discriminant features (some discriminative features have to be removed because they are irregulated), or for selecting regulated features (some irregulated features cannot be removed as they are discriminative).

Here, we present a new deep normalization model, which is based on deep generative neural nets and is designed for normalizing distributions of individual speakers. Most importantly, we will utilize the powerful normalization flow (NF) model to perform distribution transform, and propose a novel discriminative NF (DNF) model to deal with multiple speakers. To the best knowledge of the authors, this represents a new research direction in this domain.

\subsection{Normalization flow}


Deep generative models transform a simple distribution via a deep neural net, so that the output distribution matches the true data~\cite{mackay1995bayesian}.  Typical deep generative models include generative adversarial networks (GAN)~\cite{goodfellow2014generative} and variational auto-encoders (VAE)~\cite{kingma2013auto}.  Normalization flow (NF) is another deep generative model, which is similar to VAE but the transform is invertible, therefore it does not require an explicit encoder and the likelihood can be computed exactly~\cite{papamakarios2019normalizing}.  In this research, we choose NF as the basic architecture of our deep normalization model.

The foundation of NF is the principle of distribution transformation for continuous variables~\cite{rudin2006real}. Let a latent variable $\mathbf{z}$ and an observation variable $\mathbf{x}$ be linked by an invertible transform $\mathbf{x} = f(\mathbf{z})$, their probability density has the following relationship~\cite{rudin2006real}:

\begin{equation}
\label{eq:flow}
\ln p(\mathbf{x}) = \ln p(\mathbf{z}) + \ln \Big | \det \frac{ \partial f^{-1}(\mathbf{x})}{\partial \mathbf{x}} \Big |,
\end{equation}

\noindent where $f^{-1}(\mathbf{x})$ is the inverse function of $f(\mathbf{z})$. It has been shown that if $f$ is flexible enough, a simple distribution, which we assume to be a standard Gaussian, can be transformed to a very complex distribution~\cite{papamakarios2019normalizing}. Note that the second term on the right side of the above equation represents the volume (entropy) change during the transform.

\begin{figure}[htb]
    \centering
    \includegraphics[width=1\linewidth]{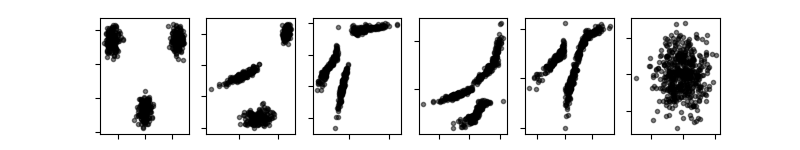}
    \caption{A complex mixture of three Gaussians is transformed to a single Gaussian by an NF model.  The pictures were generated by using two-dimensional simulation data.
    The NF used here is a masked autoregressive flow (MAF)~\cite{papamakarios2017masked}, and the distribution on $\mathbf{z}$ is a standard Gaussian.}
    \label{fig:change}
\end{figure}

Usually, $f$ is implemented as a composition of a sequence of relatively simple invertible transforms, denoted by $f_0, f_1, ..., f_T$:

\begin{equation}
f = f_T \cdot f_{T-1} ... \cdot f_0,
\end{equation}

\noindent where every $f_i$ can be a structured neural net~\cite{tabak2013family}. The entire transform has the following relationship:

\begin{equation}
\ln p(\mathbf{x}) = \ln p(\mathbf{z}) + \sum_{t=1}^{T+1} \ln \Big | \det \frac{ \partial f_{t-1}^{-1}(\mathbf{z}_t)}{\partial \mathbf{z}_t} \Big |,
\end{equation}

\noindent where we have defined $\mathbf{z}_0 = \mathbf{z}$, $\mathbf{z}_{T+1} = \mathbf{x}$ and $\mathbf{z}_{t+1} = f_t(\mathbf{z}_t)$.  This model resembles a flow of transforms, which reshapes the simple distribution on $\mathbf{z}$ gradually, and ultimately reaches the complex distribution on $\mathbf{x}$.  In the inverse direction, it normalizes the complex distribution on $\mathbf{x}$ to a simple distribution on $\mathbf{z}$, and is therefore called a \textbf{normalization flow}.  Fig.~\ref{fig:change} illustrates how a complex distribution is normalized to a simple distribution by an NF model.

The NF model can be trained with the maximum likelihood (ML) criterion. Note that Eq.\ref{eq:flow} formulates a distribution density $p(\mathbf{x})$ on the observation $\mathbf{x}$, where the first term $p(\mathbf{z})$ is often called the \textbf{prior distribution}, and the second term is called the \textbf{entropy term}. The ML training optimizes the NF model with the following objective:

\begin{eqnarray}
L(\pmb{\theta}) &=& \sum_i \ln p(\mathbf{x}_i) \nonumber \\
                &=& \sum_i \ln p(\mathbf{z}_i) + \sum_i \sum_{t=1}^{T+1} \ln \Big | \det \frac{ \partial f_{t-1}^{-1}(\mathbf{z}_{it})}{\partial \mathbf{z}_{it}} \Big |,
\end{eqnarray}

\noindent where $i$ indexes the training samples, and $\pmb{\theta}$ represents the parameters of the model. Once the model has been well trained, it can be used to (a) sample $\mathbf{x}$ by sampling $\mathbf{z}$; (b) compute $p(\mathbf{x})$ by calculating the prior $p(\mathbf{z})$ and the entropy term; (c) normalize $\mathbf{x}$ by transforming it to $\mathbf{z}$, which is Gaussian distributed. In this paper, we will focus on utilizing the normalization capability of the NF model.

The key issue when designing the NF model is to identify a net structure so that the entropy term in Eq.~{\ref{eq:flow}} can be easily computed.  Researchers have proposed various NF models based on different structures.  These models can be categorized into volume preserved (VP)~\cite{dinh2014nice} and non-volume preserved (NVP)~\cite{dinh2016density} models.  Although VPs do not change the volume during the flow (i.e. the entropy term is zero), NVPs do not have this constraint and so are generally more flexible.

\subsection{Discriminative normalization flow}

\begin{figure}[htb]
    \centering
    \includegraphics[width=1\linewidth]{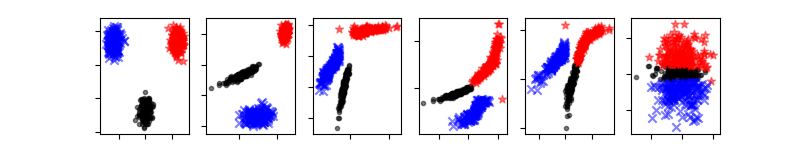}
    \includegraphics[width=1\linewidth]{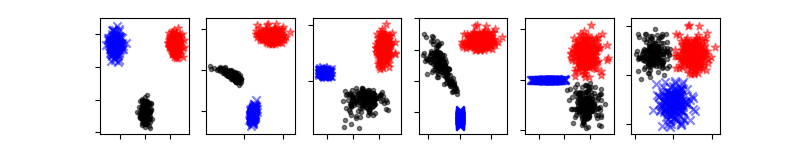}
    \caption{Vanilla NF (top) pulls all classes together in the latent space, while DNF (bottom) keeps data from different classes separated. The pictures were generated by using two-dimensional simulation data.}
    \label{fig:congress}
\end{figure}

The vanilla NF model optimizes the distribution of the training data without considering the class labels, i.e. the marginal distribution.  This means that data from different classes tend to congest together in the latent space, and the distributions of individual classes are non-Gaussian, as shown in the top row of Fig.~\ref{fig:congress}.  This is not a good property for classification tasks like speaker recognition. Conditional NF models~\cite{ardizzone2019guided} may take the class information as a condition variable, however the conditioning cannot be generalized to unseen classes (e.g. unknown speakers), which makes it unsuitable for open-set tasks such as speaker recognition.

In order to normalize distributions of individual classes and keep different classes separated, we propose a discriminative normalization flow (DNF) model.  The main difference is that we allow each class to have its own Gaussian prior, i.e. all the priors share the same covariance but possess different means, formulated as follows:

\begin{equation}
p_y(\mathbf{z}) =  N(\mathbf{z}; \pmb{\mu}_y, \pmb{\Sigma}),
\end{equation}

\noindent where $y$ is the class label. By setting class-specific means, different classes will be separated from each other in the latent space, as shown in the bottom row of Fig.~\ref{fig:congress}.

Training DNF is mostly the same as the vanilla NF, following the ML criterion. The only difference is that the probability of an observation $\mathbf{x}$ should be evaluated with the prior corresponding to its class label, formally written by:

\begin{equation}
 p(\mathbf{x}) = p_{y(\mathbf{x})}(\mathbf{z}) \Bigg|\det \frac{\partial f^{-1}(\mathbf{x})}{\partial \mathbf{x}}\Bigg|,
\end{equation}

\noindent where $y(\mathbf{x})$ is the class label of $\mathbf{x}$, and $\mathbf{z}=f^{-1}(\mathbf{x})$.  Pooling all the training data, we obtain the objective function for DNF training:

\begin{equation}
L(\pmb{\Theta}) = \sum_i \ln (p_{y(\mathbf{x_i})}(\mathbf{z_i})) + \ln \Bigg|\det \frac{\partial f^{-1}(\mathbf{x_i})}{\partial \mathbf{x_i}}\Bigg|,
\end{equation}

\noindent where $\pmb{\Theta}=\{\{\pmb{\mu}_y\}, \pmb{\Sigma}, \pmb{\theta}\}$ involves all the parameters of the model.  Note that this objective is a bit over-parameterized, as the covariance $\pmb{\Sigma}$ can be set to any values if the flow is flexible enough, e.g. in the case of NVP. We therefore manually set $\pmb{\Sigma}=\mathbf{I}$ and let the flow handle the volume change.

After training, the DNF model will establish a normalization space for $\mathbf{z}$, where the distribution $p_y(\mathbf{z})$ of every class $y$ is simply a Gaussian with covariance $\mathbf{I}$.  With this model, an observation $\mathbf{x}$ can be transformed to its latent code $\mathbf{z}$ by the inverse transform $f^{-1}(\mathbf{x})$, without knowing its class labels.  In addition, the latent codes from the same class, which may be unknown, tend to be a Gaussian. From this perspective, DNF is a nonlinear feature transform that is dedicated to within-class normalization.

\begin{figure*}[htb]
    \centering
    \includegraphics[width=1\linewidth]{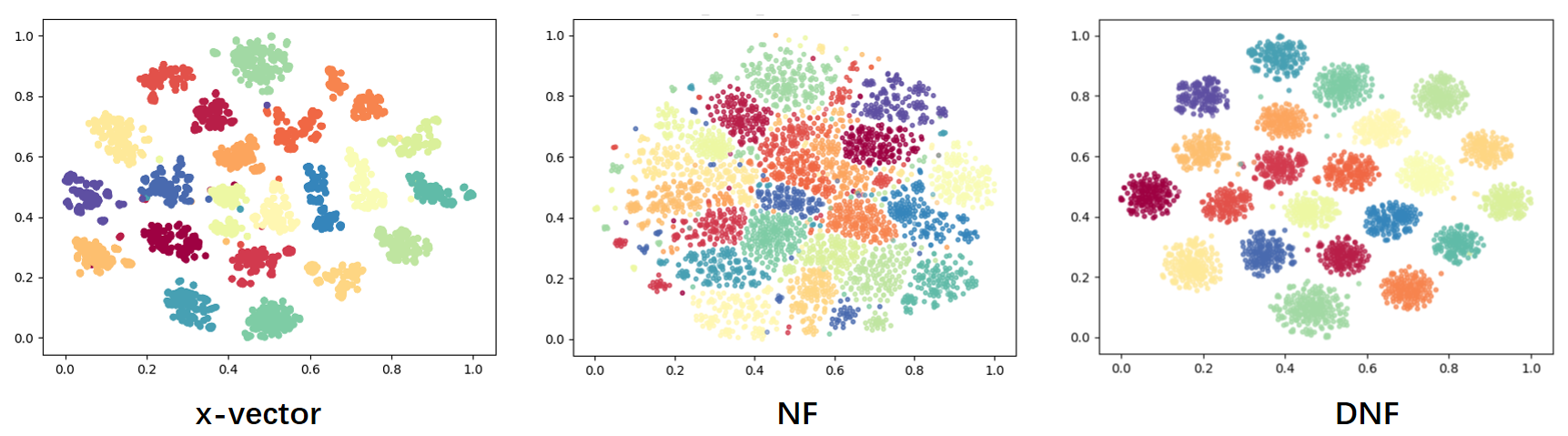}
    \caption{(Left) original x-vectors; (Middle) latent codes produced by NF; (Right) latent codes produced by DNF. Each color represents a class (speaker), and the pictures are plotted by t-SNE~\cite{saaten2008}.}
    \label{fig:dnf-real}
\end{figure*}

Fig.~\ref{fig:dnf-real} shows the distributions of x-vectors generated from the VoxCeleb dataset (refer to Section~\ref{sec:exp} for details of the generation) and the latent codes produced by NF and DNF respectively.
It is clearly to see that the class-conditional distributions become more Gaussian with DNF encoding.

\subsection{Relation to LDA}
\label{sec:dnf:lda}
From the probabilistic model view, DNF is a nonlinear extension of LDA.  Both DNF and LDA are generative models, and they share the same assumption
that the distributions of all classes are homogeneous Gaussian in the latent space. However, this assumption can never be true for LDA if the data are complex, due to the limit of the linear transform between the data space and the latent space.  However, DNF, our proposed approach, is based on a nonlinear transform, which allows it to establish a truly homogeneous and Gaussian latent space, even for complex irregulated data.

From the dimension reduction view, DNF has a similar role as the normalization step of LDA.  Both approaches normalize the distribution of data; the key difference is that the normalization step of LDA normalizes the \emph{aggregated} conditional distribution of all classes to an isotropic Gaussian, while DNF normalizes \emph{all} the conditional distributions to homogeneous isotropic Gaussians. Therefore, our DNF approach can deliver a more powerful normalization than the linear normalization of LDA.

However, it should be noted that unlike LDA, DNF does not normalize the between-class covariance, which may lead to performance loss with classification methods where dimension independence is assumed, such as those based on cosine distance. We can therefore combine DNF and LDA by substituting the linear normalization step of LDA for DNF, while keeping the linear discrimination step of LDA unchanged. This leads to a new model with a nonlinear normalization step and a linear discrimination step, which we will call \textbf{nonlinear discriminative analysis}, and we will investigate its performance in the experimental section.

\subsection{Importance of normalization}

An argument for modern speaker recognition is that when the speaker vectors become more and more discriminative, a simple back-end model based on cosine distance will be sufficient.
This argument, however, needs a careful discussion. Essentially, speaker recognition is a decision task with uncertainty, for which the optimal scores should lead to minimum Bayes risk (MBR).
Different distributions of the speaker vectors correspond to different optimal scoring models. For example, to make the cosine scoring optimal, the speaker vectors should distribute
on a spherical surface and each class distributes as a von Mises-Fisher distribution~\cite{fisher1953dispersion}. In contrast, if the data are linear Gaussian, then the
optimal scoring model is PLDA~\cite{wang2020remakes,wang2020simulation}. This means that even if the speaker vectors are sufficiently discriminative, an appropriate back-end
model is still required, according to the true distribution of the data.

Unfortunately, speaker vectors, especially those derived from deep embedding, do not follow a regularized distribution, which makes a simple back-end model not applicable.
The role of DNF plays is normalizing speaker vectors to a linear Guassian, so that PLDA can be used as an optimal back-end model. Similarly, if we can
design a model that normalizes each class to a von Mises-Fisher distribution and the class means are evenly distributed on a spherical surface, the cosine scoring will be optimal.

In summary, to improve speaker recognition performance, the speaker vectors should be discriminative \emph{and} normalized. Plenty research focuses on improving the discrimination,
but the importance of normalization has not been equally emphasized.

\section{Experiments}
\label{sec:exp}

\subsection{Datasets}

Three datasets were used in our experiments: VoxCeleb~\cite{nagrani2017voxceleb,chung2018voxceleb2}, SITW~\cite{mclaren2016speakers} and CNCeleb~\cite{fan2019cn}. VoxCeleb was used for training all the models (i-vector, x-vector, LDA, PLDA and DNF models), while the other two were used for performance evaluation.

\textbf{VoxCeleb}: This is a large-scale audiovisual speaker database collected by the University of Oxford, UK. The entire database consists of \textbf{VoxCeleb1} and \textbf{VoxCeleb2}. All the speech signals were collected from open-source media channels and therefore involve rich variations in channel, style, and ambient noise. This dataset, after removing the utterances shared by the SITW dataset, was used to train the i-vector, x-vector, LDA, PLDA and DNF models. The entire dataset contains $2000+$ hours of speech signals from $7000+$ speakers. Data augmentation was applied to improve robustness, with the MUSAN corpus~\cite{musan2015} used to generate noisy utterances, and the room impulse responses (RIRS) corpus~\cite{ko2017study} was used to generate reverberant utterances.

\textbf{SITW}: This is a standard evaluation dataset excerpted from VoxCeleb1, which consists of $299$ speakers. In our experiments, the \textbf{Eval. Core} test set, which contains $3,658$ target trials and $718,130$ imposter trials, was used for evaluation. It should be noted that the acoustic condition of SITW is similar to that of the training set VoxCeleb, so this test can be regarded as an \textbf{in-domain test}.

\textbf{CNCeleb}: This is a large-scale free speaker recognition dataset collected by Tsinghua University from source media. It contains more than $130k$ utterances from $1,000$ Chinese celebrities. It covers $11$ diverse genres, which makes speaker recognition on this dataset much more challenging than on SITW~\cite{fan2019cn}. By pair-wise composition, $107,984,700$ trials are constructed, including $119,983$ target trials and $107,864,717$ imposter trials. It is important to note that the acoustic condition of CNCeleb is quite different from that of VoxCeleb, and this therefore represents a challenging corpus that is suitable for use as an \textbf{out-of-domain test}.

\subsection{Model settings}

Our speaker recognition systems consist of three components: an x-vector or i-vector frontend that produces speaker vectors, a normalization model that regularizes the distribution of the speaker vectors, and finally, a scoring model that produces pair-wise scores for making a genuine/imposter decision.

\subsubsection{Frontend}

\begin{itemize}

\item \textbf{x-vector system}: The x-vector frontend was created using the Kaldi toolkit~\cite{povey2011kaldi}, following the SITW recipe. The acoustic features are $40$-dimensional Fbanks. The main architecture contains three components. The first component is the feature-learning component, which involves $5$ time-delay (TD) layers to learn frame-level speaker features. The slicing parameters for these $5$ TD layers are: \{$t$-$2$, $t$-$1$, $t$, $t$+$1$, $t$+$2$\}, \{$t$-$2$, $t$, $t$+$2$\}, \{$t$-$3$, $t$, $t$+$3$\}, \{$t$\}, \{$t$\}. The second component is the statistical pooling component, which computes the mean and standard deviation of the frame-level features from a speech segment. The final one is the speaker-classification component, which discriminates between different speakers. This component has $2$ full-connection (FC) layers and the size of its output is $7,185$, corresponding to the number of speakers in the training set. Once trained, the $512$-dimensional activations of the penultimate FC layer are read out as an x-vector.

\item \textbf{i-vector system}: The i-vector frontend was built with the Kaldi toolkit~\cite{povey2011kaldi}, following the SITW recipe. The raw features involve $24$-dimensional MFCCs plus the log energy, augmented by first- and second-order derivatives, resulting in a $75$-dimensional feature vector. This feature is used by the i-vector model. The universal background model (UBM) consists of $2,048$ Gaussian components, and the dimensionality of the the i-vectors is set to be $400$.

\end{itemize}

\subsubsection{Normalization models}

To investigate the merits of our proposed DNF model, we compare its performance with a number of different configurations.
Note that in all the experiments, a simple length normalization followed by whitening was employed to improve the Gaussianality of the
speaker vectors, as suggested by Garcia et al.~\cite{garcia2011analysis}.

\begin{itemize}

\item \textbf{LDA}: We implemented the basic LDA model, trained to maximize the Fisher criterion. We used the implementation in the Kaldi toolkit~\cite{povey2011kaldi}, which involves a small modification that specifies $\lambda \mathbf{S}_b + \mathbf{S}_w$ rather than $\mathbf{S}_w$, where $\lambda$ is a hyperparameter that was set to be 0.1 in the LDA + cosine scoring experiment, and 0.0 in the LDA + PLDA scoring experiment.

\item \textbf{LDA/N}: The linear normalization component of LDA. It simply normalizes the within-speaker covariance to be an identity matrix, neither diagonalizing the between-speaker covariance nor reducing any dimensions.

\item \textbf{DNF}: We implemented the proposed DNF model with the Masked Autoregressive Flow (MAF) architecture~\cite{papamakarios2017masked}.
    It consists of 10 MAF blocks, and each block is an inverse autoregressive transformation $u^i_{j}= (u^{i-1}_{j} - \mu^i_j)\exp(-\alpha^i_j)$,
    where $u^i_j$ is the $j$-th output of the $i$-th block, $\mu^i_j = f_{\mu}(u^{i-1}_{1:j-1})$ and $\alpha^i_j=f_{\alpha}(u^{i-1}_{1:j-1})$.
    $\{f_{\mu}, f_{\alpha}\}$ are unconstrained functions and implemented by three-layer full-connected neural networks in our experiment.
    The model was trained following the ML criterion with the Adam optimizer~\cite{kingma2014adam},
    where the minibatch size was set to $300$ and the learning rate was set to $0.003$.
    For more details of the MAF architecture, please refer to~\cite{papamakarios2017masked} and the source code is publicly available\footnote{https://github.com/Caiyq2019/DNF}.

\item \textbf{DNF-LDA}: One potential issue with DNF is that it does not normalize the between-class covariance. Here, we perform an additional LDA after DNF normalization, to achieve normalization on both within- and between-class covariance.  This is essentially a simple implementation of the nonlinear discriminant model discussed in Section~\ref{sec:dnf:lda}.

\end{itemize}

\subsubsection{Scoring model}

Two commonly used scoring models were applied in this study: the simple \textbf{Cosine scoring}, which is based on the cosine distance, and the more complicated \textbf{PLDA scoring}, which is based on PLDA~\cite{Ioffe06}.

\subsection{Basic results}

In the first experiment, we apply the four normalization models (LDA, LDA/N, DNF, DNF-LDA) to regulate the standard x-vectors derived from both SITW and CNCeleb. The results in terms of equal error rate (EER) are reported in Table~\ref{tab:res-x}.

\subsubsection{X-vector in-domain results}

Firstly, focusing on SITW, the in-domain test, it can be seen that all the normalization models provide performance improvement with cosine scoring. The fact that LDA/N outperforms the baseline in a very significant way (9.19 vs. 17.20) demonstrates the importance of within-speaker normalization, although this approach is only linear. DNF performs better than LDA/N (8.53 vs 9.19), confirming that nonlinear normalization is better than a linear one. LDA performs much better than LDA/N and DNF, demonstrating the importance of the between-class information.  Finally, DNF-LDA achieves the best performance, by combining the strength of DNF and LDA,.

For PLDA scoring, all the non-linear normalization models (including dimension-trunking LDA, DNF and DNF + LDA) offer performance improvement. Note that any linear transform (LDA/N and LDA without dimension reduction) does not change the PLDA performance, as the within-speaker and between-speaker covariances that PLDA relies on are linearly invariant. LDA with dimension-reduction provides reasonable performance improvement when the dimension size is carefully selected, demonstrating the importance of distribution normalization for individual speakers. Significantly, DNF obtains better performance than LDA, which confirms that NF is a better normalization approach for this problem. By adding additional LDA-based normalization, DNF-LDA achieves the best performance. These results are consistent with those obtained with cosine scoring.

\subsubsection{X-vector out-of-domain results}

When using CNCeleb, the out-of-domain test, the observations are very different.  Looking at the cosine scoring results, firstly we observe that LDA/N does not offer any performance improvement over the baseline (16.36 vs 16.32), which indicates that the global within-speaker covariances are significantly different between VoxCeleb (the training data) and CNCeleb, and so the linear normalization that is learned to diagonalize the within-speaker covariance of VoxCeleb can never diagonalize the within-speaker covariance of CNCeleb.  LDA, which applies additional transform and dimension selection based on the between-class variance, makes things even worse. This suggests that the between-class covariances of VoxCeleb and CNCeleb are also significantly different.

DNF, which focuses on normalizing distributions of individual speakers, is more robust against data mismatching, when compared to the global linear normalization with LDA/N (14.22 vs 16.36).  Applying additional LDA after DNF reduces the performance, which suggests that in the DNF latent space, the between-class covariance still changes significantly from VoxCeleb to CNCeleb. The only exception is the case of dimension-preserving DNF-LDA [512], which does not perform any dimension reduction and provides a marginal gain over DNF (13.83 vs 14.22). This indicates that in the DNF latent space, although the \emph{shape} of the between-speaker covariance has changed significantly from VoxCeleb to CNCeleb, the \emph{principle directions} of the covariance may not change much.  This is not the case within the LDA/N latent space, as the performance with LDA [512] is worse than with LDA/N (16.87 vs 16.36).

For PLDA scoring, similar conclusions can be drawn: LDA fails in most situations. The principle role of LDA in this scenario is removing irregulated dimensions, and this removal is based on the between-speaker covariance within the latent space by LDA/N, which is in turn based on the within-speaker covariance. However, as we have discussed, both the between- and within-speaker covariances change significantly from VoxCeleb to CNCeleb, so it is not surprising that the LDA-based normalization fails. In contrast to LDA, DNF still works in this situation, which can be attributed to the more robust within-class normalization. However, when applying additional LDA, the unreliable between-class information is used for dimension reduction, which leads to significant performance reduction. This is shown in the case of DNF-LDA with reduced dimensions.

To summarize, the experimental results presented above indicate that the global properties (within- and between-speaker covariances) may change significantly at the dataset level, and any normalization methods based on these properties will suffer from the generalization problem. DNF learns how to normalize individual speakers at different locations of the speaker space, which appears to be more generalizable to unseen data. However, this generalizability seems to only be for within-class distributions: after DNF normalization, there is still significant mismatch with regard to between-class distributions, which should be further investigated.

\subsubsection{I-vector results}

For the purpose of comparison, we report the results with i-vectors in Table~\ref{tab:res-i}.  It can be seen that the normalization methods make very little  contribution to PLDA scoring on both SITW and CNCeleb databases. For the cosine scoring, LDA contributes with performance gains on SITW.  We argue that this is mainly due to the diagonalization on the between-speaker covariance. However, this contribution is largely lost on CNCeleb, indicating that the between-speaker covariance has changed significantly from SITW to CNCeleb. This observation is the same as in the x-vector experiment.  DNF does not show any advantage in this experiment. This is because the within-speaker distributions of i-vectors have been well regulated (see Table~\ref{tab:var}, so a dedicated normalization is not necessary.

\vspace{-2mm}
\begin{table}[htb!]
 \begin{center}
  \caption{EER(\%) results on SITW and CNCeleb with x-vector frontend.}
  \label{tab:res-x}
   \begin{tabular}{|l|c|c|c|c|}
   \hline
                & \multicolumn{2}{c|}{SITW}   & \multicolumn{2}{c|} {CNCeleb}\\
   \hline
                & Cosine    & PLDA  & Cosine  & PLDA \\
   \hline
   x-vector [512]     & 17.20    & 5.30   & 16.32       & 13.03\\
   \hline
   LDA [150]          & 5.25     & 4.07   & 17.67       & 14.37\\
   LDA [200]          & 5.82     & 3.96   & 17.52       & 13.50\\
   LDA [400]          & 7.38     & 4.65   & 17.49       & 12.28\\
   LDA [512]          & 8.61     & 5.30   & 16.87       & 13.03\\
   LDA/N [512]        & 9.19     & 5.30   & 16.36       & 13.03 \\
   \hline
   DNF [512]          & 8.53     & 3.66   & 14.22       & \textbf{11.82}\\
   \hline
   DNF-LDA [150]    & \textbf{5.06}  & 3.61   & 15.42       & 13.85\\
   DNF-LDA [200]      & 5.41     & \textbf{3.42}  & 15.18       & 13.22\\
   DNF-LDA [400]      & 7.05     & 3.58   & 14.20       & 11.90\\
   DNF-LDA [512]      & 8.17     & 3.66   & \textbf{13.83}       & \textbf{11.82}\\
   \hline
  \end{tabular}
 \end{center}
\end{table}
\vspace{-5mm}

\begin{table}[htb!]
 \begin{center}
  \caption{EER(\%) results on SITW and CNCeleb with i-vector frontend.}
  \label{tab:res-i}
      \begin{tabular}{|l|c|c|c|c|}
   \hline
                & \multicolumn{2}{c|}{SITW}   & \multicolumn{2}{c|} {CNCeleb}\\
   \hline
                      & Cosine       & PLDA      & Cosine      & PLDA \\
   \hline
   i-vector [400]     & 14.24        & 5.66     & 17.68       & 18.25\\
   \hline
   LDA [150]          & \textbf{7.11}  & 5.36     & 18.18       & 18.49\\
   LDA [200]          & 7.46         & \textbf{5.25} & 17.85       & 18.36\\
   LDA [400]          & 9.32         & 5.66     & \textbf{16.65}  & 18.25\\
   LDA/N [400]        & 11.84        & 5.66     & 17.23       & 18.25\\
   \hline
   DNF [400]          & 12.06        & 5.60     & 18.04       & \textbf{18.15}\\
   \hline
   DNF-LDA [150]      & 7.30         & 5.41     & 18.30       & 18.53\\
   DNF-LDA [200]      & 7.52         & 5.30     & 18.02       & 18.36\\
   DNF-LDA [400]      & 9.49         & 5.60     & 17.02       & \textbf{18.15}\\
   \hline
   \end{tabular}
  \end{center}
\end{table}
\vspace{-5mm}

\subsection{Results on more powerful x-vectors}

In this experiment, we constructed more powerful x-vector systems to investigate whether DNF normalization still contributes. We conducted extensive preliminary trials on model structures and training objectives (not reported here due to space constraints), and based on these, we chose three architectures to represent SOTA systems.

\begin{itemize}
\item  {\bf TDNN + Att.}: The same architecture as the TDNN baseline in the previous experiment, but the statistical pooling is replaced by self-attention pooling~\cite{zhu2018self}.
\item  {\bf ResNet-34 + Att.}: The ResNet-34 architecture~\cite{chung2018voxceleb2,zeinali2019but} with self-attention pooling~\cite{zhu2018self}.
\item  {\bf ResNet-34 + AAM}: The ResNet-34 architecture~\cite{chung2018voxceleb2,zeinali2019but} with additive angular marginal loss~\cite{deng2019arcface}.
\end{itemize}

The experimental results are shown in Table~\ref{tab:res-xx}. For LDA, we report the LDA [200] results only, as it is the best configuration for all the
LDA systems. On SITW, all these `advanced' systems outperform the TDNN baseline in a significant way, and DNF still achieves good performance. In most situations, DNF outperforms LDA, and more performance gains can be attained by DNF-LDA. The results on CNCeleb are even more significant.  Firstly, they once again confirm the generalizability of DNF, as reported previously; and secondly, they show that the EER reduction on SITW provided by the `advanced' approaches was not transferred to the results on CNCeleb.  This indicates that the performance improvement obtained with some of the `advanced' techniques may simply be the result of overfitting.

\begin{table}[htb!]
 \begin{center}
  \caption{EER(\%) results on SITW and CNCeleb with a SOTA x-vector frontend.}
  \label{tab:res-xx}
   \begin{tabular}{|l|l|c|c|c|c|}
   \hline
                    & & \multicolumn{2}{c|}{SITW}   & \multicolumn{2}{c|} {CNCeleb}\\
   \hline
                    &                    & Cosine    & PLDA   & Cosine  & PLDA \\
   \hline
   TDNN             & x-vector [512]     & 17.20     & 5.30   & 16.32   & 13.03\\
                    & LDA [200]          & 5.82      & 3.96   & 17.52   & 13.50\\
                    & DNF [512]          & 8.53      & 3.66   & 14.22   & 11.82\\
                    & DNF-LDA [200]      & 5.41      & 3.42   & 15.18   & 13.22\\
   \hline
   TDNN + Att.      & x-vector [512]     & 4.37      & 3.66   & 15.08   & 13.05 \\
                    & LDA [200]          & 3.72      & 2.73   & 18.34   & 13.97 \\
                    &  DNF [512]         & 5.00      & 2.71   & 14.69   & 12.07\\
                    &  DNF-LDA [200]     & 3.72      & 2.57   & 15.45   & 13.66\\
  \hline
   ResNet-34 + Att.  & x-vector [512]    &  2.73     & 2.52   & 13.94   & 13.11\\
                     & LDA [200]         &  2.60     & 2.00   & 14.90   & 12.58\\
                     & DNF [512]         &  3.47     & 1.94   & \textbf{13.86} & \textbf{11.61}\\
                     & DNF-LDA [200]     &  \textbf{2.57} & 1.89   & 14.04   & 12.32\\
  \hline
   ResNet-34 + AAM  &x-vector [512]      &  5.71    &  2.82    & 15.80  & 14.02 \\
                    &LDA [200]           &  2.73    &  1.86    & 16.67  & 13.42 \\
                    &DNF [512]           &  4.89    &  2.32    & 14.66  & 12.80\\
                    &DNF-LDA [200]       &  2.93    &  \textbf{1.83} & 14.96  & 12.59\\
   \hline
  \end{tabular}
 \end{center}
\end{table}
\vspace{-5mm}

\subsection{Analysis}
\label{sec:analysis}

To better understand the behavior of DNF, we monitored the training process, and here we report the change of the statistics related to regulation and discrimination. As in Section~\ref{sec:lda} when analyzing LDA, we conduct the analysis with a small-scale experiment. All the data and measures are the same as in the LDA investigation, and we focus on x-vector results.



\subsubsection{Regulation analysis}

\begin{figure}[htb]
    \centering
    \includegraphics[width=\linewidth]{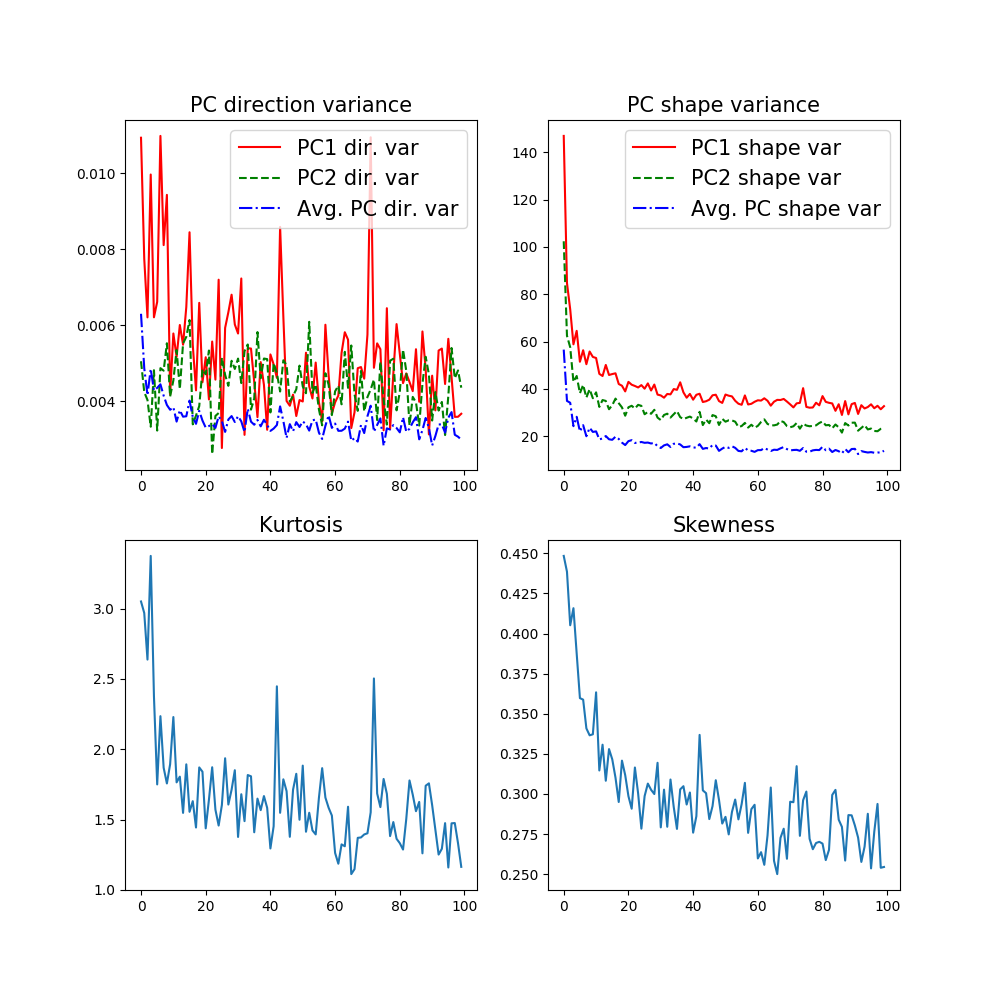}
    \vspace{-5mm}
    \caption{Change of measures related to data homogenity and Gaussianality during DNF training.}
    \label{fig:norm}
\end{figure}

Fig.~\ref{fig:norm} presents the four groups of measures related to data regulation: PC directional variance (PC dir. var) and PC shape variance (PC shape var), which reflect the homogenity of distributions of different speakers, and averaged kurtosis and averaged skewness, which reflect the Gaussianality of the distributions of each speaker. All these measures have been used in Section~\ref{sec:metric}
and the exact meaning can be found there.

It can be seen that the values of all these measures are significantly reduced during training. Compared to the results in Table~\ref{tab:var}, it can be seen that the DNF can generally reach lower values on all these measures compared to LDA, hence is a better normalization model. Spikes are found with kurtosis, skewness, and PC1/PC2 direction variance.  These spikes indicate that the model is trying to change the location of all the speakers in order to find an optimal configuration, but changing one speaker may cause unwanted change on other speakers, due to the complex distributions of the speaker vectors. Nevertheless, the training can ultimately find a better configuration that improves the data regulation in general.

\subsubsection{Discrimination analysis}
\begin{figure}[htb]
    \centering
    \includegraphics[width=\linewidth]{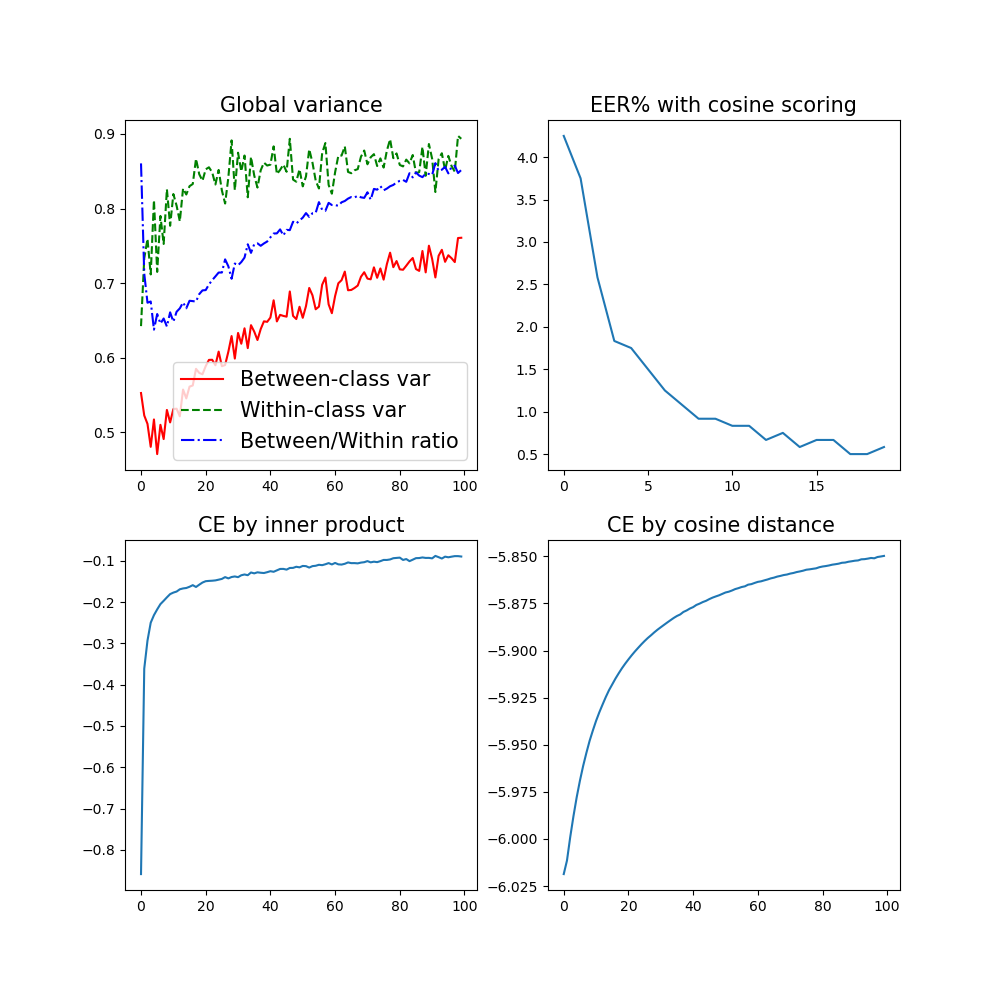}
    \vspace{-5mm}
    \caption{Change of measures related to class discrimination during DNF training. }
    \label{fig:disc}
\end{figure}

To investigate the discriminative capability of the DNF-normalized speaker vectors, we compute several measures related to class discrimination: (1) between-class and within-class variance and their ratio; (2) EER results based on cosine scoring; (3) cross entropy (CE) between the predicted class posterior and the true one-hot class label, i.e., $\sum_n \log [\sigma (y)]_c$, where $c$ is the true class label, $\sigma$ is the softmax function. The logit $y$ (activation before softmax) is computed as the inner product of training samples and all the class means; (4) cross entropy as in (3), but the logit $y$ is computed as the cosine distance between training samples and the class means. Fig.~\ref{fig:disc} shows the change of these measures during model training. This shows that the data in the latent space becomes increasingly discriminative over time, as indicated by all these measures. In particular, we highlight the continuous increase of the cross entropy based on inner product.  If we treat the inverse NF function $f^{-1}(\mathbf{x})$ as a regular neural net and the mean vectors of all the classes as the weights of the final layer, the whole DNF architecture is a standard classification network. This net is usually trained with the CE loss. In DNF, we interpreted the net in a very different way (a generative model) and trained it with a very different loss (ML), and obtained the same CE reduction. This confirms the fundamental relation between generative and discriminative models, as discussed in Section~\ref{sec:dnf}.

\section{Conclusions}
\label{sec:con}

This paper investigated the issue of data irregulation with deep speaker vectors in speaker recognition, and found through comprehensive experiments that deep speaker vectors require deep normalization. Firstly, we found that the within-speaker distributions of deep speaker vectors are highly non-homogeneous and non-Gaussian, which may seriously impact performance of speaker recognition systems. To overcome this problem, we introduced a new deep normalization approach, based on a novel discriminative normalization flow (DNF) model. This model is a nonlinear extension of LDA, and can normalize complex and heterogeneous distributions of individual speakers. Using the state of the art system configurations, our experiments on two datasets demonstrated that our new DNF approach delivers consistently better performance compared to the baseline and outperforms the more conventional LDA-based normalization. Furthermore, in the out-of-domain test where LDA performs very poorly, DNF still delivers good performance, confirming the good generalizability and further potential of our approach.  Future work will investigate the joint training of the DNF normalizer and the speaker embedding model, and will also apply DNFs to raw acoustic features directly.

\
\bibliographystyle{IEEEtran}
\bibliography{mybib}

\vspace{-5mm}
\begin{IEEEbiographynophoto}{Yunqi Cai} studied as a phd candidate at Institute of Physics ,Beijing, from 2014 to 2018. During 2016-2017 he went to Oak ridge national lab, the United States, as a visiting scholar under the funding of the Joint PhD Training Program scholarship offered by UCAS 2016. He received his Ph.D. degree from Institute of Physics, Chinese Academy of Sciences, Beijing, in 2018. He is now a Postdoctoral Fellow with the Center for Speech and Language Technologies (CSLT) and the Department of Computer Science at Tsinghua University, Beijing, China. His research interests are focused on the machine learning algorithms of speech and language technology.
\end{IEEEbiographynophoto}

\vspace{-5mm}
\begin{IEEEbiographynophoto}{Lantian Li} received the B.Sc. degree from China University of Mining and Technology, Beijing in 2013. He received the Ph.D. degree from the Department of Computer Science, Tsinghua University in 2018. Since 2018, he has been with the Center for Speech and Language Technology (CSLT), Tsinghua University as a postdoctoral fellow. His research interest is speaker recognition with machine learning methods.
\end{IEEEbiographynophoto}

\vspace{-5mm}
\begin{IEEEbiographynophoto}{Dong Wang}(M'09, SM'16) received the B.Sc. and M.Sc. degrees in computer science from Tsinghua University in 1999 and 2002. He received the Ph.D. degree (supported by a Marie Curie fellowship) from CSTR, University of Edinburgh, in 2010. He was employed with Oracle China during 2002-2004 and IBM China during 2004-2006. He joined CSTR, University of Edinburgh, in 2006 as a Research Fellow. From 2010 to 2011, he was with EURECOM as a Postdoctoral Fellow, and from 2011 to 2012, was a Senior Research Scientist with Nuance. He is now an Associate Professor with Tsinghua University, Beijing, China.
\end{IEEEbiographynophoto}

\vspace{-5mm}
\begin{IEEEbiographynophoto}{Andrew Abel} is a lecturer in Computing Science at Xi'an Jiaotong-Liverpool University (XJTLU) in Suzhou, China, and received his Ph.D. from the University of Stirling in Scotland in 2013, after conducting research into developing signal-image processing algorithms for enabling multi-modal speech processing technologies. Before XJTLU, he worked as a researcher at Stirling, working on the testing of novel MEMS/CMOS microphone technology, at Anhui University in Hefei focusing on image processing, and more recently at Stirling as a research-CI  on the development of next generation hearing aid technology aimed at developing hearing aids that can ``see''.  His research interests are focused on the use of multiple modalities (particularly vision) to estimate and process speech and communication.
\end{IEEEbiographynophoto}

\end{document}